\documentclass[pre,twocolumn,nofootinbib]{revtex4}
\usepackage{amssymb,amsmath,amsthm}
\usepackage[dvips]{graphicx}
\usepackage{verbatim}

\newcommand{\be}{\begin{equation}}
\newcommand{\ee}{\end{equation}}
\newcommand{\bea}{\begin{eqnarray}}
\newcommand{\eea}{\end{eqnarray}}

\begin{document}

\title{Entropy correlation distance method applied to study correlations between the Gross Domestic Product of rich countries } 
\author{Marcel Ausloos$^*$, Janusz Mi\'s{}kiewicz$^\dag$}
\affiliation{$^*$GRAPES, ULg., B5a, B-4000 Li$\grave e$ge, Euroland\\
$^\dag$Institute of Theoretical Physics, University of Wroc\l{}aw, pl. M.
Borna 9, 50-204 Wroc\l{}aw, Poland}

\date{Jan 15, 2009 }

\begin{abstract}

\noindent The Theil index is much used in economy and finance; it looks like the Shannon entropy, but pertains to event values rather than to their probabilities. Any time series can be remapped through the Theil index. Correlation coefficients can be evaluated between the new time series,  thereby allowing to study their mutual statistical distance, - to be contrasted to the usual correlation distance measure for the primary time series.  As an example this entropy-like correlation distance method (ECDM) is applied to the Gross Domestic Product of 20 rich countries in order to test some economy globalization process. Hierarchical distances allow to construct (i) a linear network, (ii) a Locally Minimal Spanning Tree. The role of time averaging in finite size windows is illustrated and discussed. It is also shown that the mean distance between the most developed countries, was decreasing since 1960  till 2000, - which we consider to be a proof of globalization of the economy for these countries. 
 
\bigskip

\end{abstract}

\maketitle

\bigskip

\noindent {\bf 1. Introduction} \smallskip

\noindent The Theil\footnote{H. Theil was a Dutch econometrician who was born on 13 October 1924 in Amsterdam, graduated from the University of Amsterdam, succeeded to Jan Tinbergen at the Erasmus University Rotterdam, moved and taught later in Chicago and at the University of Florida. He died in 2000.} index, often used in economy and finance, is defined through
\be
Th(x;N) = \frac{1}{N} {\sum_{i=1}^{N} \left(\frac{x_i}{\langle x \rangle}  \ln \frac{x_i}{\langle x \rangle}\right)}.
\label{eq:defTh}
\ee
It served to measure the distribution of income $x_i$ of  $i$ agents, among $N$ agents, with respect to the average income $\langle x  \rangle$, --  the average being taken over the ensemble of incomes of the population of size $N$. $Th(x;N)$ spans the range $0$ till ln($N$). $x$ is the vector of data $(x_1,\ldots , x_N)$.
It looks like the Shannon entropy but was invented to consider the event values themselves rather than their probability of occurrence. One peculiarity is that it measures the agent's share relative to the mean $\langle x \rangle$ of the population. In terms of information theory, the Theil index measures the difference between the maximum entropy and its present value. An interesting development is to consider that the $x_i$ quantity in Eq.(\ref{eq:defTh}) is time dependent. Thus one can generalize the Theil index in order to remap a time series $x(t)$ in a nonlinear way into a $Th(t)$, as done in Sect. 2 which recalls considerations outlined in [Mi\'skiewicz, 2008]. Thereafter from the Theil mapped series one can look at time dependent correlations between different data sets, distances, hierarchies, and other usual features, through various techniques of data analysis, like those leading to or resulting from network constructions. 

The first application is here below made to macroeconomy time series, in particular to the GDP of 20 among the  richest countries. Following up on studies of correlations between GDPs of rich countries [Mi\'skiewicz \& Ausloos, 2005; Mi\'skiewicz \& Ausloos, 2007; Ausloos \& Lambiotte,  2007; Ausloos \& Mi\'skiewicz, 2008; Mi\'skiewicz, 2008; Mi\'skiewicz \& Ausloos, 2008; Ausloos \& Gligor,  2008; Gligor \& Ausloos,  2008a;  Gligor \& Ausloos,  2008b], we have analyzed web-downloaded data on GDP\footnote{from the Conference Board and Groningen Growth and Development Centre, Total Economy Database, September 2008, $http://www.conference-board.org/economics$}, used as individual wealth signatures of a country economical state (``status''). We have calculated the fluctuations of the Theil mapped GDP in different time windows and looked for correlations, and subsequent distances, as reported in Sect. 2. 

Usually, a complex system can be represented by a network, -- nodes being scalar, i.e. agents, here the countries, while links are weights, here measures of distances between two $Th(t)$ taken from GDP fluctuation correlations between two countries. Indeed time series can be represented by networks [Yang \& Yang, 2008].
In order to extract structures from the networks, we have also averaged the time correlations in different windows.  This allows more robustness in the subsequent networks properties and reveals  evolving {\it statistical  distances}. 
In line with our previous work we have examined two different (so called) networks.
The results are presented in Sect. 3. A brief discussion on economy globalization follows with a conclusion in Sect. 4. It is found that such a measure of {\it collective habits} does fit the usual expectations defined by politicians or economists,  i.e. common factors are to be searched for.

\bigskip

\noindent {\bf 2. From macroeconomy index input to network construction} \smallskip

\noindent {\bf 2.1. GDP data} \smallskip

\noindent GDP data sets of  several among the most rich OECD countries were used for illustrating the method, i.e. 20 countries: Austria (AT), Belgium (BE), Canada (CA), Denmark (DK), Finland (FI), France (FR), Greece (GR), Ireland (IR), Italy (IT), Japan (JP), the Netherlands (NL), Norway (NO), Portugal (PT), Spain (ES), Sweden (SE), Switzerland (CH), Turkey (TK), U.K. (UK), U.S.A (US), and Germany (DE), allowing for a linear superposition of the data before the reunification in 1991 in the latter case; an ``ALL'' country is also invented as in previous works [Mi\'skiewicz \& Ausloos, 2005; Mi\'skiewicz \& Ausloos, 2006; Mi\'skiewicz \& Ausloos, 2008; Mi\'skiewicz, 2008] as a sum of the GDP of considered countries.\footnote{The set deviates somewhat from previous works [ Mi\'skiewicz \& Ausloos, 2006; Mi\'skiewicz \& Ausloos, 2008; Mi\'skiewicz, 2008] since there is neither Iceland nor Luxembourg but there is Turkey in the present paper.} Thus there are $N=21$ time series to examine. 

\begin{figure}
\includegraphics[bb=50 50 230 176]{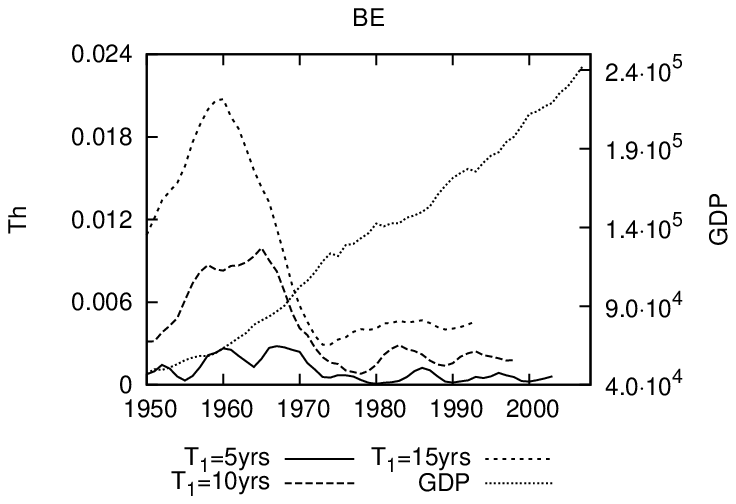} 
\vspace{0.5cm}

\includegraphics[bb=50 50 230 176]{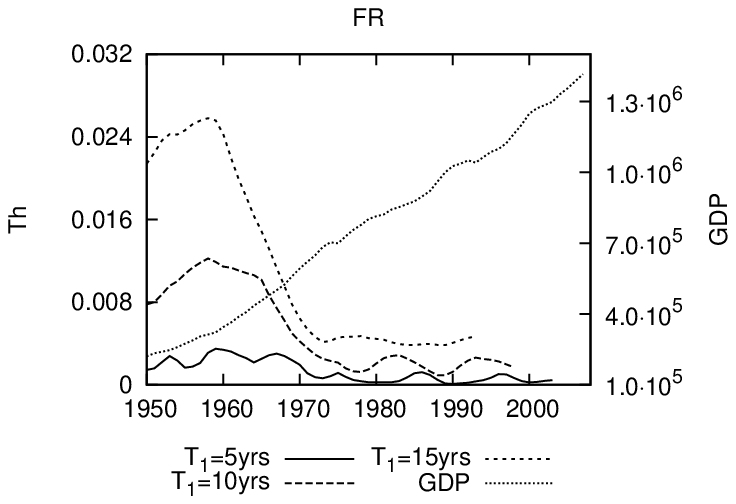}

\vspace{0.5cm}
\includegraphics[bb=50 50 230 176]{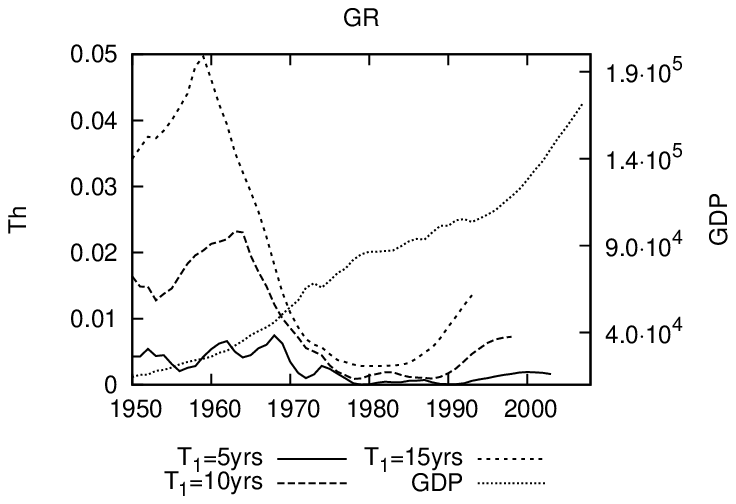} 

\vspace{0.5cm}
\includegraphics[bb=50 50 230 176]{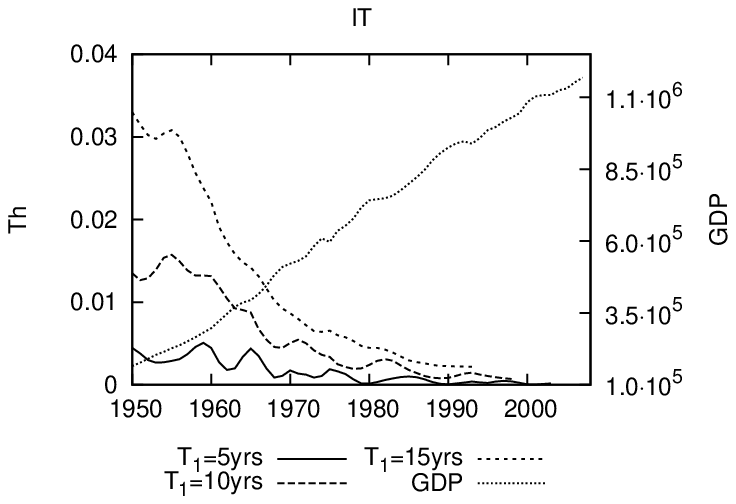}
\caption{(right y-axis) Raw GDP time series for BE, FR, GR, IT in milions of US dollars, and  (left y-axis) the resulting mapping into a Theil index for three time windows: $T_1$= 5, 10, 15 yrs. Years on the   bottom x-axis correspond to the initial  point of the time interval  }
\label{fig:theil_GDP_EUR_1}
\end{figure}

\begin{figure}
\includegraphics[bb=50 50 230 176]{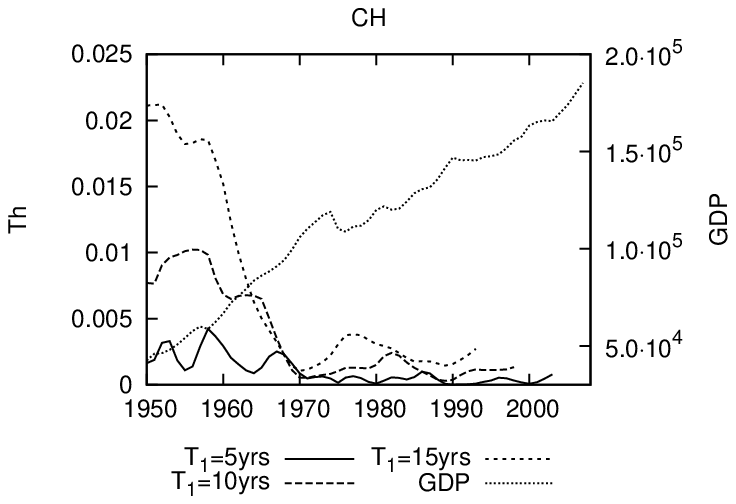} 

\vspace{0.5cm}
\includegraphics[bb=50 50 230 176]{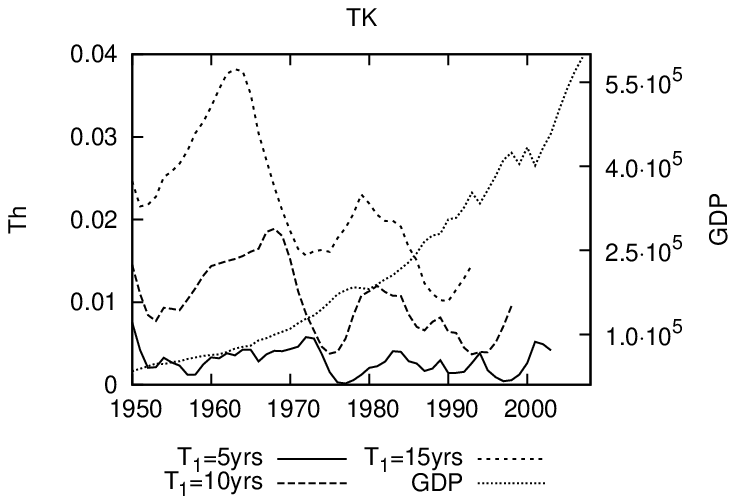}

\vspace{0.5cm}
\includegraphics[bb=50 50 230 176]{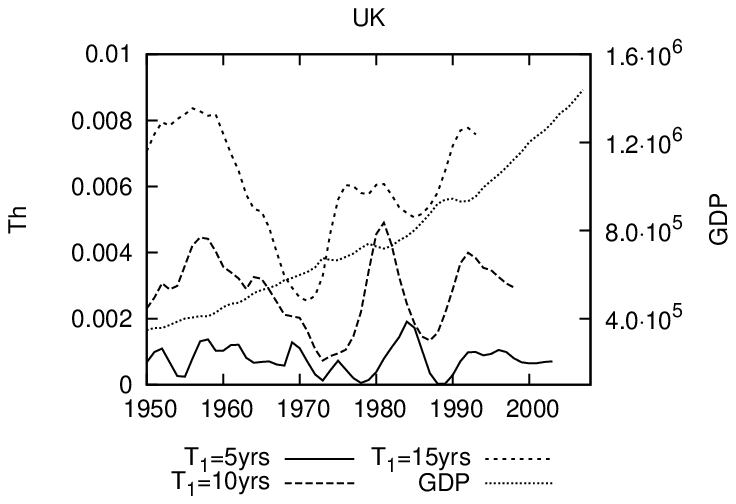} 

\vspace{0.5cm}
\includegraphics[bb=50 50 230 176]{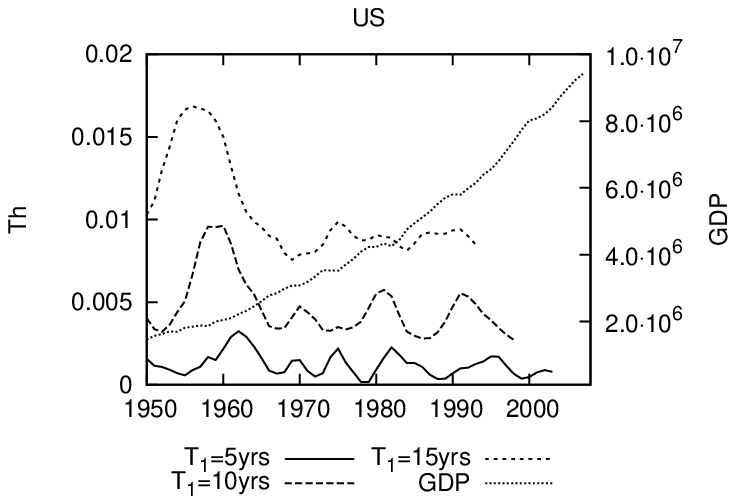}
\caption{ (right y-axis) Raw GDP time series for CH, TK, UK, US in milions of US dollars, and (left y-axis) the resulting mapping into a Theil index for three time windows: $T_1$= 5, 10, 15 yrs. Years on the  bottom x-axis correspond to the initial   point of the time interval  } \label{fig:theil_GDP_EUR_2}
\end{figure}

The data are taken from  Groningen Growth and Development Centre
($http://www.ggdc.net/index-dseries.html$). The GDP's are presented in milions of 1990 US dollars (converted at Gery Khamis PPPs). In our case  each time series starts in 1950 and finishes in 2007, such that there are 58 data points in every time series. 
The evolution of a few cases are shown in Figs. \ref{fig:theil_GDP_EUR_1}-\ref{fig:theil_GDP_EUR_2}. The GDP values range between $14 \cdot 10^{10}$ \$ and $171 \cdot 10^{10}$ \$ in the case of GR up to $1.4 \cdot 10^{12}$ \$ and $9.4 \cdot 10^{12}$ \$ for USA . Except some small ``perturbations'' the GDP of all presented countries is growing in time. Some deviation from the monotonical grow can be observed e.g. in the case of CH in 1973-1976 or in TK in 1998-2000.

\bigskip

\noindent {\bf 2.2. Mapping onto the Theil index} \smallskip

\noindent The Theil index  $Th$ can be used to nonlinearly map an original time series $A(t)$ into a new one through 
\be
Th_A(t,T_1) = \frac{1}{T_1} {\sum_{i=t}^{t+T_1} \left(\frac{A_i}{\langle A \rangle_{(t,T_1)}}  \ln \frac{A_i}{\langle A \rangle_{(t,T_1)}}\right)}
\label{eq:th}
\ee
where the average  $\langle A \rangle_{(t,T_1)} $ is made over the ensemble of points $j$ in a time window of size $T_1$, placed between $t$ and $t+T_1$:
\be
{\langle A \rangle_{(t,T_1)}} =\frac{1}{T_1} {\sum_{j=t}^{t+T_1} A_j},
\label{eq:mean}
\ee
i.e. the Theil index is calculated for the interval $[t,t+T_1]$. Here  $A_i(t)$ is the GDP of  country $i$. 

Characteristics maximum Theil index values as a function of the time window size are given in Table \ref{tab:Theilindexmaxrank}.  These $Th$ values vary between 0 and 0.05. Recall that $Th$ is 0 if all $A_i = \langle A \rangle_{(t,T_1)}  $ and is maximum if for one $i$, $A_i = N \langle A \rangle_{(t,T_1)} $ and all other $A_j$ =0. In Table \ref{tab:Theilindexmaxrank} we have emphasized the 17 European countries which we examined. It is remarkable that there are a few but weak variations in the ranking, as a function of the window size. UK has  always the smallest $Th$. The Scandinavian countries come next, followed by the western countries, and finally the mediterranean ones. But DE is always the last one having a large $Th$. It is remarkable that IR is a ... mediterranean country. This class might be rather labelled maritime. Finally, let us observe that $Th$ for $T_1=15$ yrs seems to lead to the most intuitive (geography and economy based) grouping.

\begin{table}
\begin{tabular}{|c|c|c|c|c|c|c|c|c|c|c| } \hline

$T_{1}$ & UK& NO & DK& SE &BE &FR &NL & CH& FI& IT \\ \hline
5&191 & 240 & 271 &276 & 280 & 350 & 422 & 425 & 438 & 508 \\  \hline 
$T_{1}$ &PT& AT &GR& TK& IR& ES&DE&US&CA&JP\\ \hline
5& 567 & 693 & 748& 756 & 846 & 1077&1598&324&370&1264 \\ \hline \hline
$T_{1}$ & UK& SE& NO& DK&BE &CH &FI & NL& FR& IT \\ \hline
10 &490 & 859 &867 &968&995  & 1023 & 1101 & 1167 & 1224 & 1573 \\  \hline 
$T_{1}$ & AT& PT &TK& GR& ES& IR&DE&US&CA&JP \\ \hline
10 & 1601 & 1879 & 1888 & 2324 &2724& 2865 & 2932&963&1194&3997 \\  \hline \hline
$T_{1}$ & UK& SE & NO& DK&BE &FI &CH&NL& FR& AT \\ \hline
15&837 & 1667 & 1758 &1902 &2074 &2085 & 2119 &2388 &2582 &2837 \\  \hline
$T_{1}$ & IT& TK &PT& IR&GR& ES&DE&US&CA&JP \\ \hline
15 & 3301 & 3819 &3832&4813 & 4966 &5061&5183&1686&2366&8386 \\ \hline
\end{tabular}
\caption{Maximum value of the Theil index ($\cdot 10^{5}$) calculated during the time interval of interest, for three time windows   $T_1$=5, 10, 15 yrs. The countries are ranked in decreasing $Th$ value. US, CA, and JP have been extracted from the list and placed at the end of the data display in order to emphasize the European hierarchy. }
\label{tab:Theilindexmaxrank}
\end{table}

For illustration, we give in Figs. \ref{fig:theil_GDP_EUR_1}-\ref{fig:theil_GDP_EUR_2}, the Theil index mapping of a few significant GDP time series, in particular for the following countries, in (BE, FR, GR, IT) and outside (CH, TK, UK, US)  the EUR zone and for a few $T_1$ time windows, i.e. 5, 10, 15 yrs . 

The GDP Theil index of the presented countries for the medium ($T_1=10 \; yrs$) and long ($T_1=15 \; yrs$) size time windows has a maximum at {\it ca.} 1960. In the case of the shortest time window $T_1=5 \; yrs$ the time evolution of the GDP Theil index remains on the same level without any visually meaningful extremal point, which makes such a behaviour difficult to analyse beyond a trivial statement.

A few observations can be made. In the case of BE and FR, on one hand, and CH, on the other hand, the Theil index is decreasing from 1960 till 1970 and  remains on a stable level thereafter (specially for the long time window). The Theil index of US besides the main maximum at 1960 for the medium size time window ($T_1 = 10 \; yrs$) has two other local maxima, i.e. at 1980 and 1990. However for the longest presented time window ($T_1=15 \; yrs$) only the main maximum at 1960 can be distinguished; after 1970 a relatively stable value of GDP Theil index can be observed, -- as was pointed out, the same as in BE, FR and CH. The Theil index of TK has its main maximum {\it ca.} 1970, which is followed by a decrease till a minimum at 1975 and an increase until a maximum at 1980.

For the last few data points the Theil index of TK is increasing. A similar situation can be observed in the GR case, -- the Theil index is increasing since 1985. IT seems to have more pronounced oscillations than the three other illustrating EUR countries, but the main maximum seems to occur much earlier that for the other countries.
 
The Theil index of UK for the medium size time window ($T_1=10 \; yrs$) has two pronounced maxima, one at 1960 and the second at 1980; an analogous behaviour can be made for the 15 yrs time window, but the first maximum is at 1955 and the second 1965. Finally US seems after the pronounced maximum in the early part of the data have reached a stable level on which are superposed marked oscillations.

\bigskip 

\noindent {\bf 2.3. Time series distances} \smallskip

\noindent In order to compare  time series, one can measure characteristic features, like their Hurst exponent, their (multifractal and power) spectrum, .... or their relative distance. Several definitions of distances can be found in the literature. The distance between two time series (here the Theil-mapped time series) is hereby defined as the absolute value of the difference between mean values in the interval $[t,t+T_2]$. For further considerations to be explained below, in Sec. 4, one could also consider non-equal time correlations, thus taking into account a time lag $\tau$. Therefore we define
\be
d_{Th}(A,B)_{(t,T_1,T_2,\tau)} = \vert \langle Th^A(t,T_1) - Th^B (t+\tau,T_1)  \rangle_{(t,T_2)} \vert .
\label{eq:th_1}
\ee 
In Eq.(\ref{eq:th_1}) the mean value denoted by brackets ($\langle ... \rangle $) is defined as in Eq.(\ref{eq:mean}). Such a mean value can be taken on a time window $T_2$ different from $T_1$. In the present paper we will only report and discuss data when $\tau=0$. As a result we have two time parameters:
\begin{enumerate}
\item the $T_1$ remapping time window while calculating the $Th$ index and
\item the correlation window $T_2$.
\end{enumerate}

\noindent {\bf 2.4. Network  construction} \smallskip

\noindent The distance between nodes matrices obtained from Eq.(\ref{eq:th_1}) are here below analysed after constructing two network structures and measuring their statistical properties. The following networks are considered: (i) the unidirectional minimal length path (UMLP) and (ii) the locally minimal spanning tree (LMST). The algorithms generating the mentioned networks are:
\begin{description}
\item[UMLP]
The network begins with an arbitrary chosen country, -- here the ALL country, then the closest neighbouring country is attached and become the new end of the network. The next country closest to this end of the network is searched and attached. The process continued until all countries are attached.
\item[LMST]
The root of the network is the pair of closest neighbouring countries. Then the country closest to any node is searched and attached. The algorithm is continued until all countries are attached to the network.
The ``ALL'' country is not used in the network construction.
 \end{description}

We emphasize that in the UMLP construction, ALL is at the beginning of the chain, while in the other two constructions, ALL is treated as an ordinary country. The LMST network seed is the appropriate pair of the closest countries according to the appropriate distance matrix. The first network is linear, and essentially robust against a perturbation, like removing or adding a country or in the case of a regrettable mathematical error, since it is based on a measure relative to a statistical mean, while the LMST is obviously a tree, -  rather compact when only 21 data points, thus with very few branching levels, are involved in the construction. It is known that such a tree is far from robust.

Since two time windows (Theil mapping and correlation measure) are used simultanousely the total size of the time window is equal to the sum of the time windows. In our analysis UMLP and LMST networks were constructed for all time windows ranging from $T_{1}=5$ to $58$ yrs , $T_2=1$ y  to $58$ yrs, moving along the time axis by a one year step. Therefore the number of the generated UMLP and LMST networks is equal to the time series length minus the total time window size, i.e. the $T_1$ and $T_2$ parameters must satisfy the inequality $T_1+ T_2 \leq 58$ yrs,  whence the number of generated networks ($Net$) depends on the time window sizes and is equal to $\nu=58-T_1-T_2$.

\bigskip

\noindent {\bf 3. Results} \smallskip

\bigskip

\noindent {\bf 3.1. Theil index  distance statistics} \smallskip

\begin{figure}
 \centering
\includegraphics[bb=50 50 266 201]{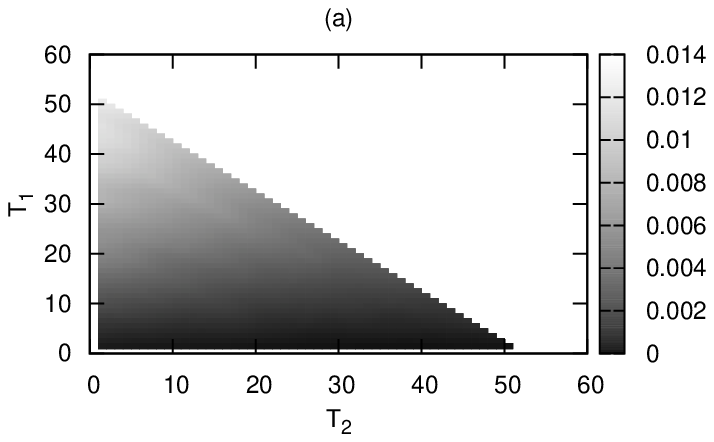}
 \includegraphics[bb=50 50 266 201]{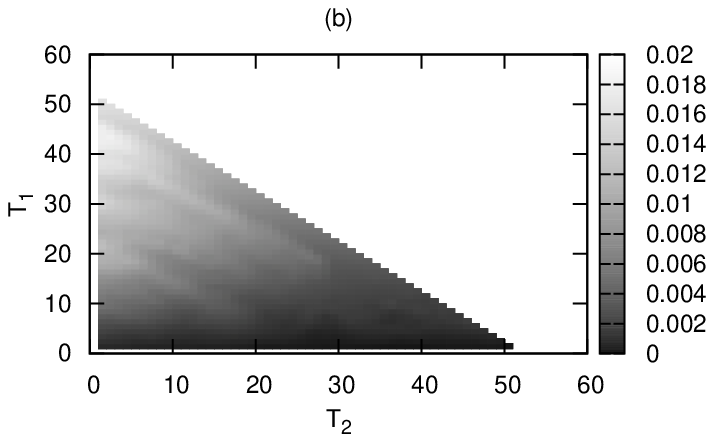}
\caption{ (a) Mean distance  and (b) standard deviation of the distance distribution between countries in a UMLP networks as a function of the $T_1$ and $T_2$ time window sizes. The distance and standard deviation result from averaging over the network links and networks generated in the moving time window.  }
\label{fig:theil_umlp_m}
\end{figure}

\begin{figure}
\centering
\includegraphics[bb=50 50 266 201]{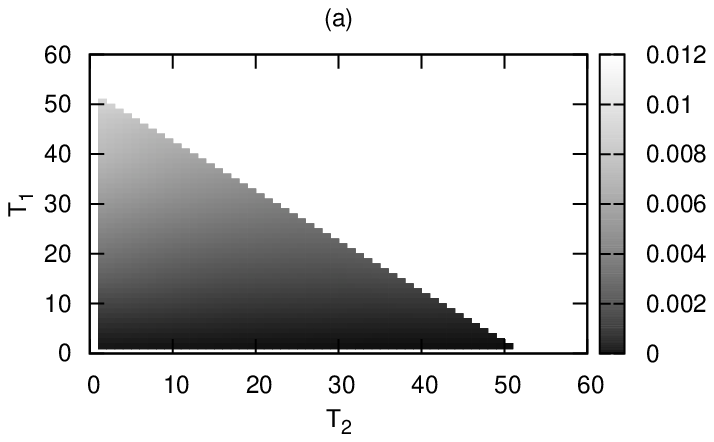}
 \includegraphics[bb=50 50 266 201]{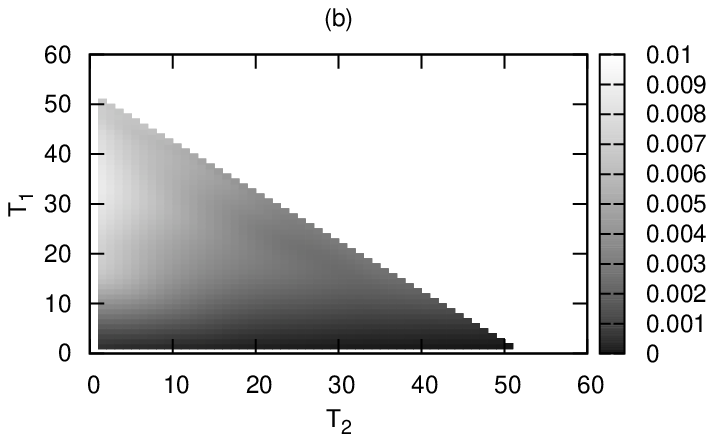}
\caption{(a) Mean distance and (b) standard deviation of the distance distribution between countries in a LMST networks as a function of the $T_1$ and $T_2$ time window sizes. The distance is averaged over the network links and networks generated in the moving time window. }
 \label{fig:theil_lmst_m}
\end{figure}

\noindent In total there is a huge number of networks. Therefore some cases are to be extracted for the present report.\footnote{All cases are available from the authors upon request.} We propose a visualisation of the data through a spectrogram method, using for the $x$ and $y$ axis respectively the time window $T_2$ and $T_1$. The data values are represented by a grey pixel in a convenient order as indicated in the figures \ref{fig:theil_umlp_m}, \ref{fig:theil_lmst_m}.

\begin{table}
\begin{tabular}{|l|l|l|l|l|l|l|l|l|} \hline
 & \multicolumn{8}{|c|}{mean $\cdot 10^{4}$} \\ \cline{2-9}
& max & $T_{1}$ & $T_2$ & $\nu$&min & $T_{1}$ & $T_2$ & $\nu$ \\ \hline
 UMLP & 130 & 56 & 1 &1 & 1.96 & 5 & 39 & 14 \\  \hline
LMST & 97  & 56 & 1 & 1 & 1.45 &5&51  & 2  \\  \hline
& \multicolumn{8}{c|}{std $\cdot 10^{4}$}\\ \cline{2-9}
& max & $T_{1}$ & $T_2$ & $\nu$&min & $T_{1}$ & $T_2$ & $\nu$\\ \hline
 UMLP &  190 & 49 & 1 & 8 & 3.29 & 5 & 36 & 17\\  \hline
LMST & 89.8  & 38 & 1 & 19 &1.92  & 5&52&1 \\  \hline
\end{tabular}
\caption{Upper part of the Table: Maximum and minimum mean Theil distance value of each type of network UMLP and LMST. The mean value is calculated over the distances between nodes on the network and the ensemble generated for the given time windows paremeters. The values of the averaging windows $T_1$ and $T_2$ when this maximum (minimum) occurs are indicated; the corresponding number of networks $\nu$ is indicated. Bottom part of the Table (std) indicates the maximum and minimum values of the std for the parameter cases so indicated.}
\label{tab:Theil}
\end{table}

The mean value and the standard deviation of the distances between nodes as a function of the $ T_1$ and $T_2$ are presented in Figs. \ref{fig:theil_umlp_m} - \ref{fig:theil_lmst_m} for UMLP and LMST cases respectively.
The largest value of the mean distance, the minimum mean distance, the maximum and minimum standard deviations as a function of the time windows $T_1$, $T_2$ are presented in Table \ref{tab:Theil}.

It can be first generally observed that the mean distance between countries and the corresponding standard deviation are the biggest for UMLP networks and the smallest for LMST networks. The maximum of the mean distance occurs for the longest $T_1$ and the shortest $T_2$ windows sizes. The minimum mean distance is found with the opposite combination of the time windows sizes, i.e. small $T_1$ and large $T_2$. 
Again let it be emphasized that the (max or min)  mean values do {\it NOT NECESSARILY } occur at  the (max or min) standard deviations.

\bigskip

\noindent {\bf 3.2. Theil network evolution} \smallskip

\begin{figure}
 \centering
 \includegraphics[bb=50 50 410 302,scale=0.5]{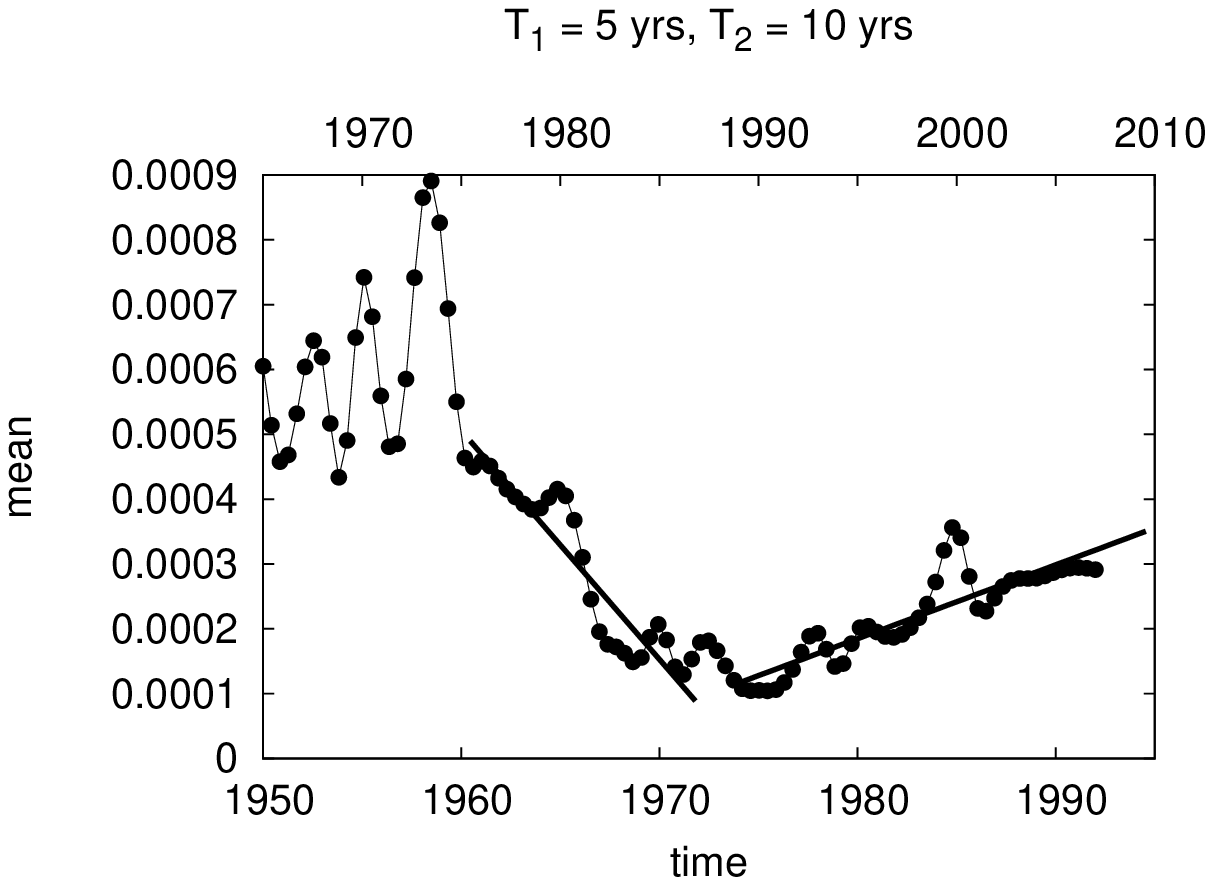}
\includegraphics[bb=50 50 410 302,scale=0.5]{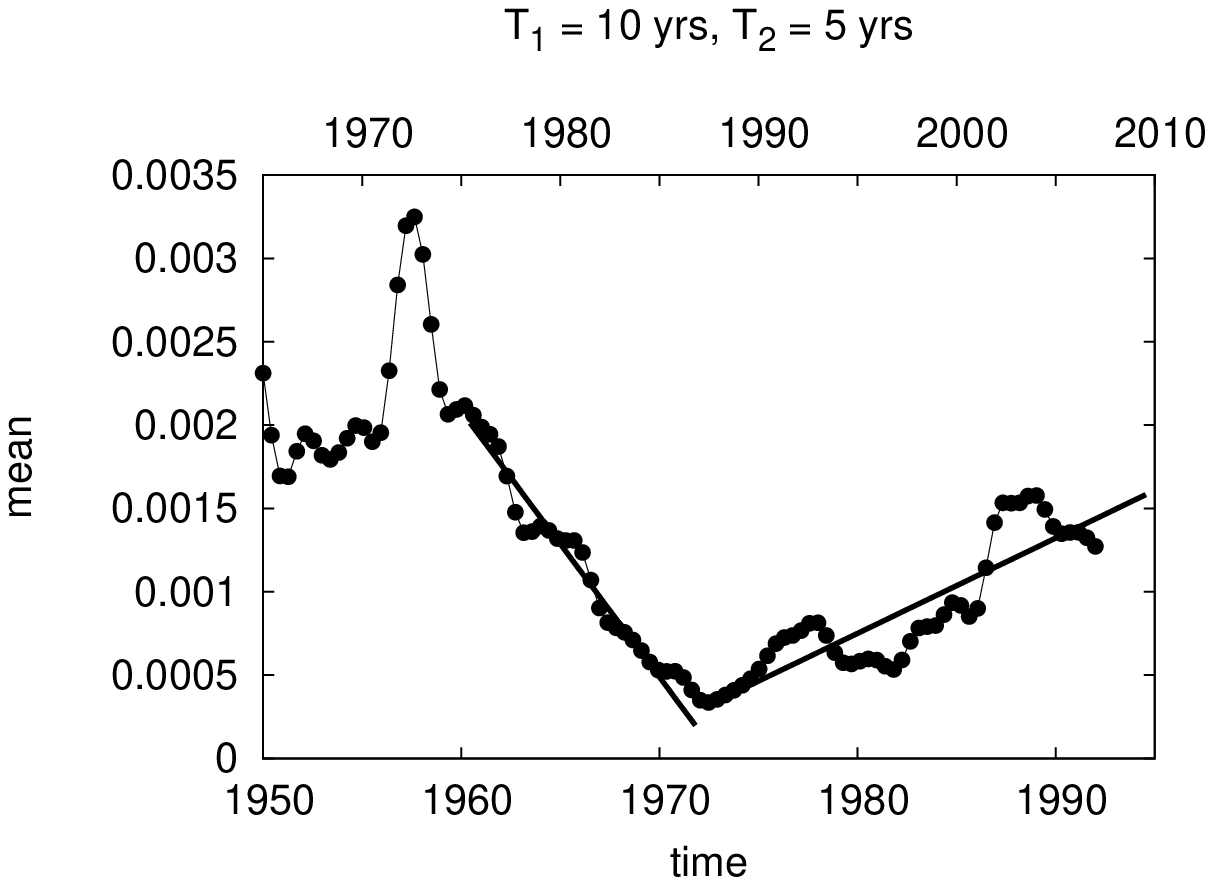}
\includegraphics[bb=50 50 410 302,scale=0.5]{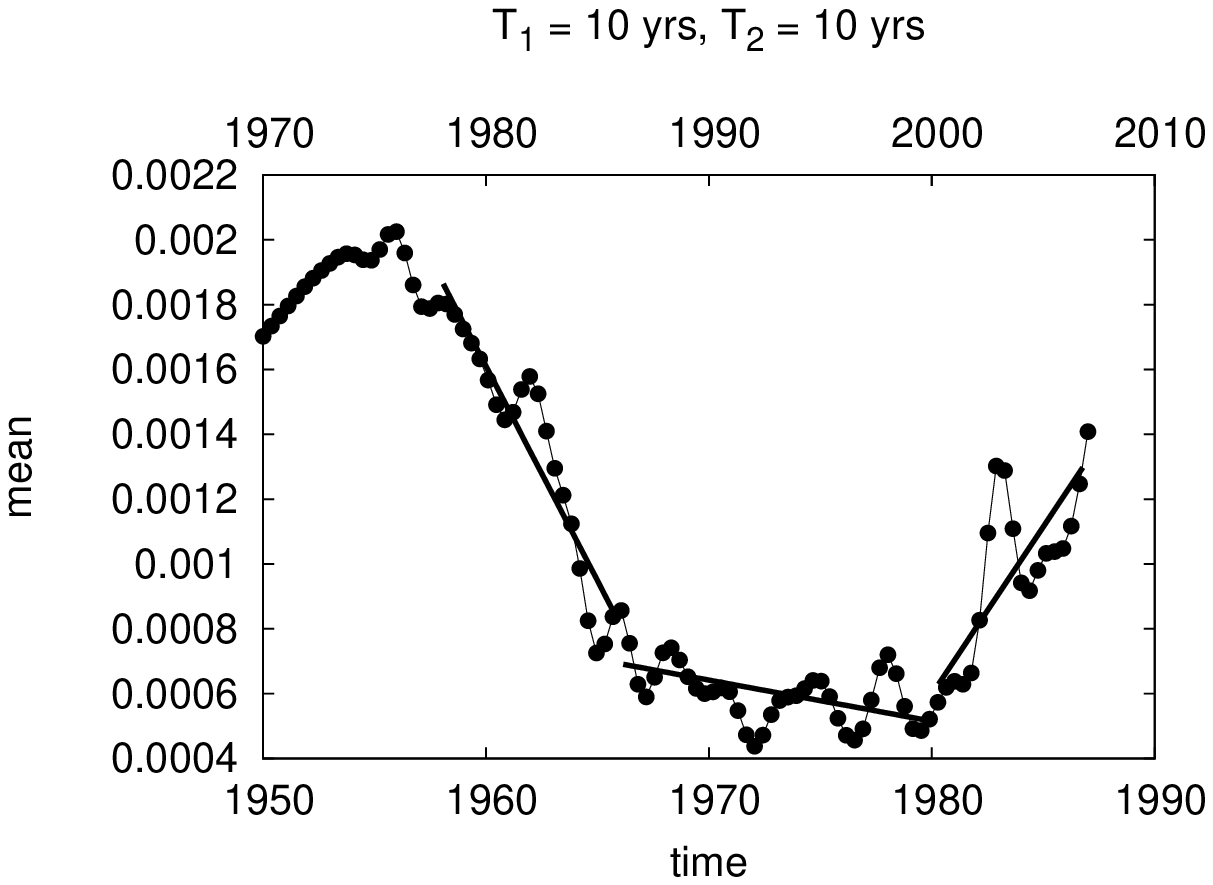}
\includegraphics[bb=50 50 410 302,scale=0.5]{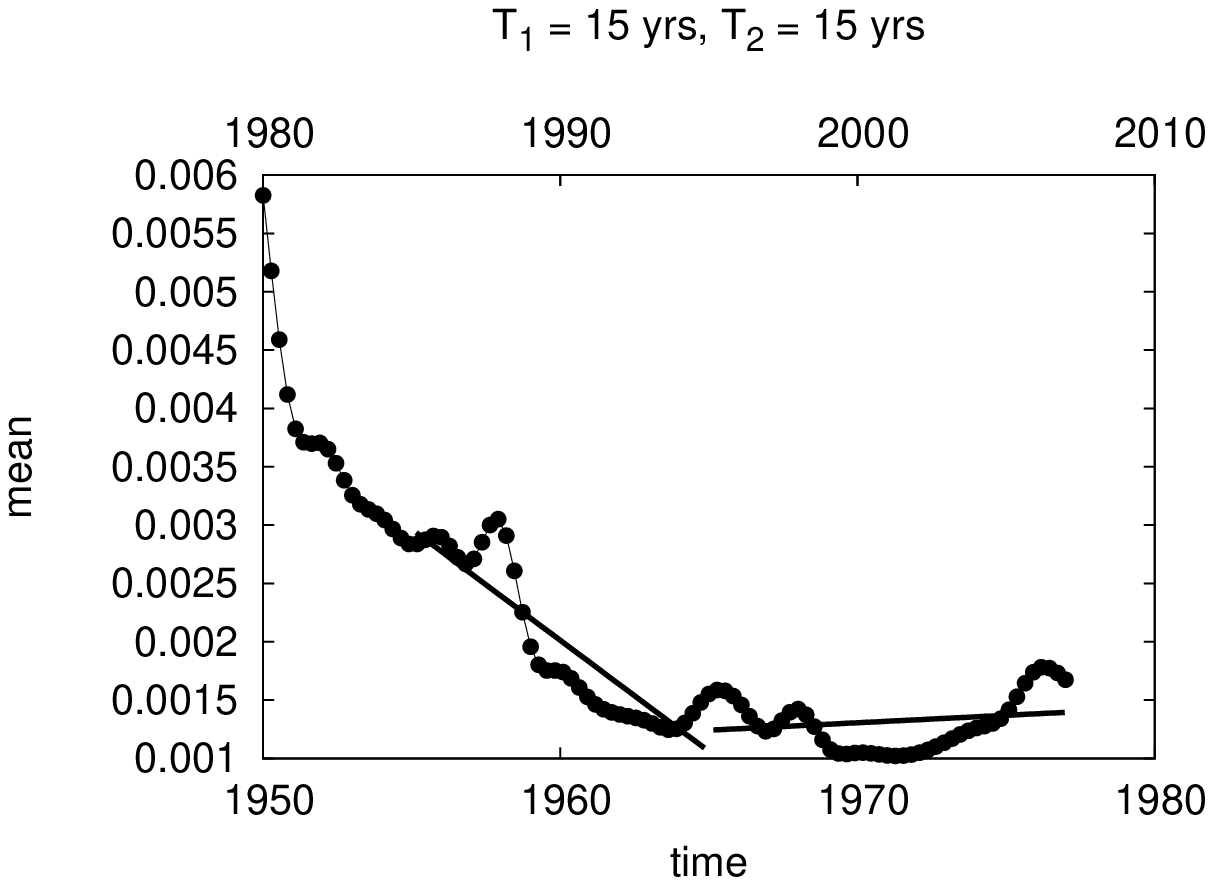}
 \caption{Yearly evolution of the mean value of the links (=distances) between nodes for the UMLP network deduced from Theil mapping analysis of GDP of the 20 examined countries. The time window sizes are given above every plot.
}
 \label{fig:tsallis_umlp_hist}
\end{figure}

\begin{figure}
 \centering
\includegraphics[bb=50 50 410 302,scale=0.5]{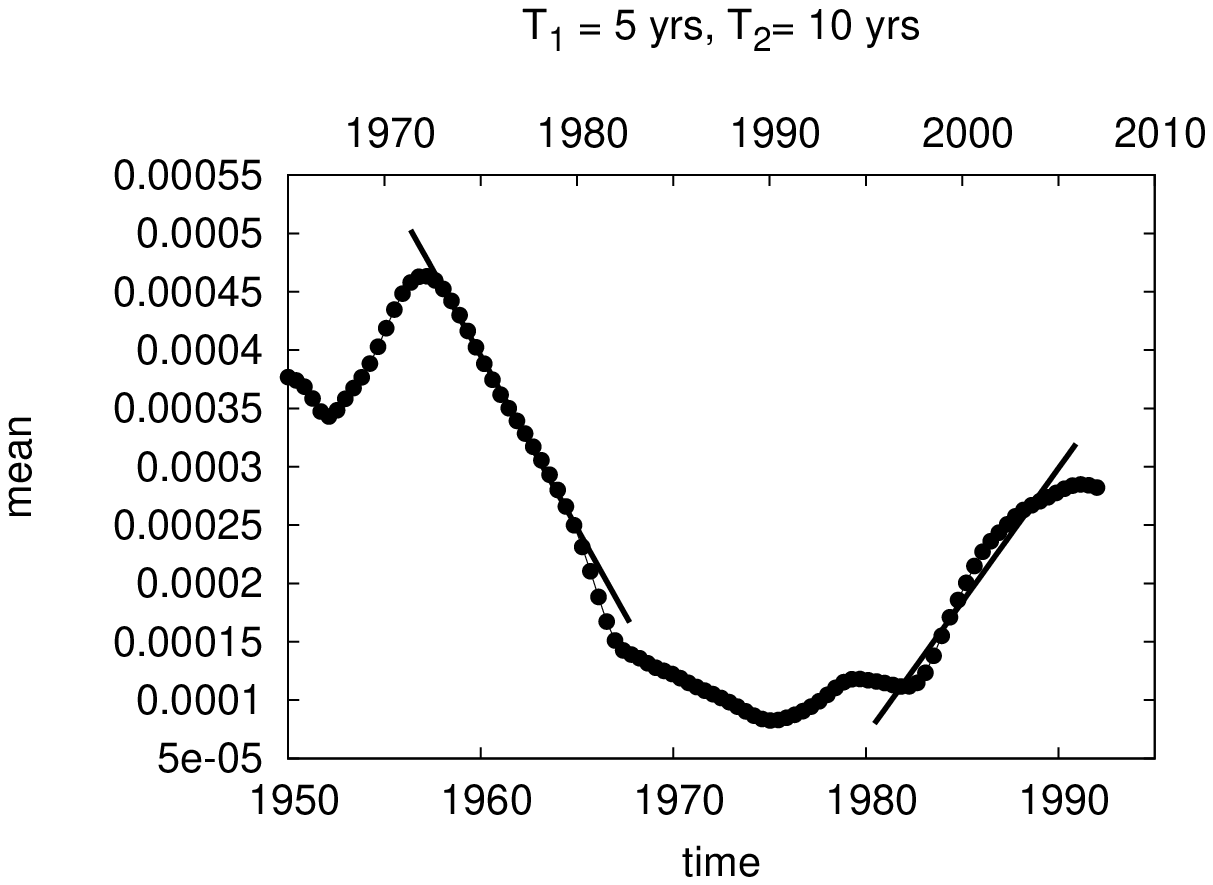}
\includegraphics[bb=50 50 410 302,scale=0.5]{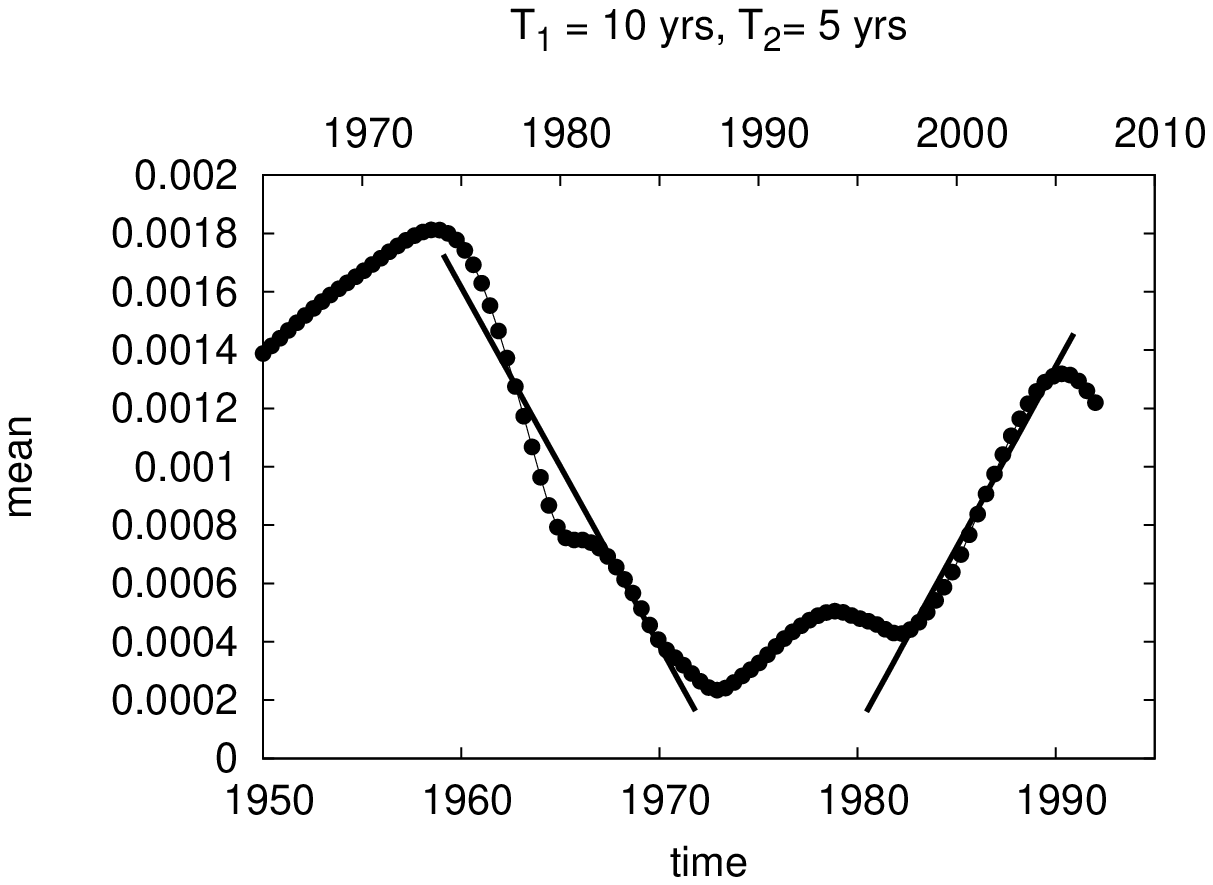}
\includegraphics[bb=50 50 410 302,scale=0.5]{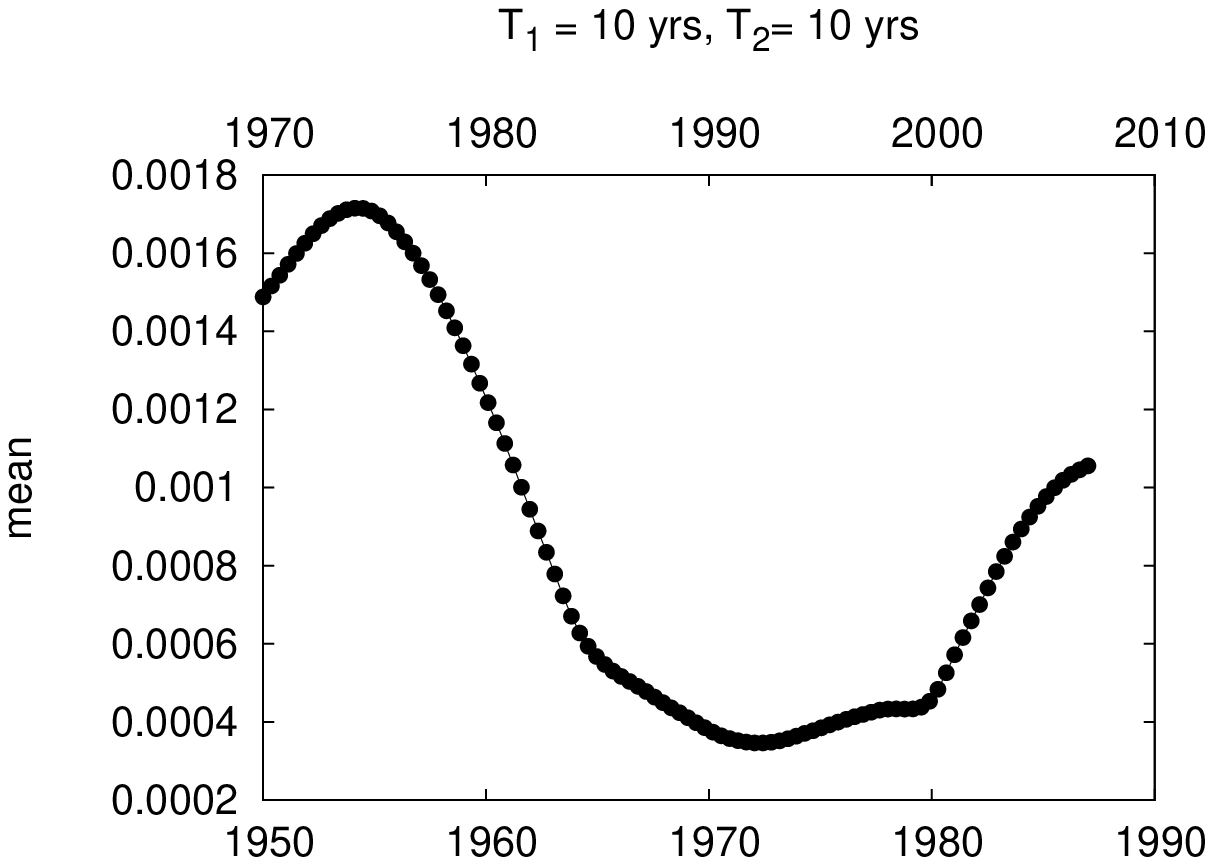}
\includegraphics[bb=50 50 410 302,scale=0.5]{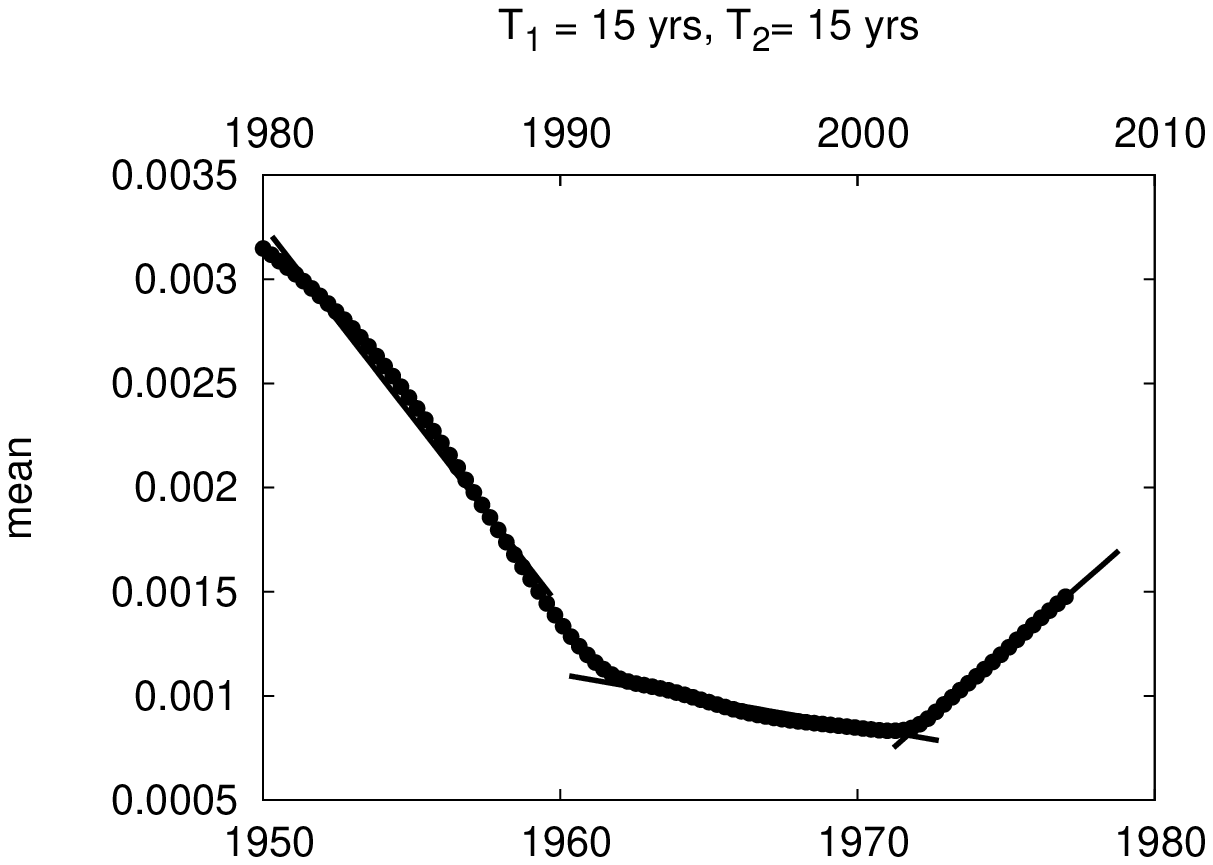}
 \caption{Yearly evolution of the mean value of the links  (=distances) between nodes for the LMST network deduced from Theil mapping analysis of GDP of the 20 examined countries. The time window sizes are given above every plot.
}
\label{fig:tsallis_lmst_hist}
\end{figure}

\noindent For further discussion the  following  time window size combinations were selected, i.e. $(T_{1}=5$ yrs, $T_2=10$ yrs$)$, $(T_{1}=10$ yrs, $T_2=5$ yrs$)$, $(T_{1}=10$ yrs, $ T_2=10$ yrs$)$, $(T_{1}=15$ yrs, $ T_2=15$ yrs$)$. The evolutions of the mean distance between countries are presented in Figs. \ref{fig:tsallis_umlp_hist}-\ref{fig:tsallis_lmst_hist}.  Straight lines indicate visually remarkable features.

The general observations to be made at this stage are the following ones:
\begin{itemize}
\item In all considered networks (UMLP and LMST) and for all window sizes three types of evolution can be distinguished: increase, decrease and relatively stable mean distances between countries. 
\item The ratio max/min of the mean networks size for the considered time windows span between 6 and 13.
\item It is worth noticing that for time windows [$(T_{1}=5 $ yrs, $T_2=10$ yrs$)$, $(T_{1}=10$ yrs, $ T_2=5 $ yrs$)$, $(T_{1}=10 $ yrs, $ T_2 =10 $ yrs$)$] the maximum of the mean distance occurs at about 1960, and 
\item since then the size of the network(s) is fast decreasing over a decade up to 1970. 
\item Thereafter the mean distance remains small and relatively stable up to 2000 or so.
\item The mean size is reincreasing after 2000.
\end{itemize}


\bigskip

\noindent {\bf 4. Conclusions} \smallskip

\noindent In conclusion, the most interesting results of this analysis are
\begin{itemize}
\item The $Th$ index values are quite small indicating a rather homogeneous set of values for the GDP centered around the mean.
\item UK has the lowest $Th$ index, indicating the most stable development.
\item ES and DE have the largest $Th$ index which might be a result of political perturbations of this latter country in the investigated time interval.
\item $Th$ values can be surprisingly grouped according to climatic rather than geographical regions..
\item Long time window size of Theil index and short correlation window size results in bigger network size and higher standard deviation of the distance distribution. Such network might be suitable for clique formation analysis.
\item The 15 yrs time window seems to give the most coherence from the intuitive grouping point of view.
\item A low value of  $Th$ does not necessarily mean that the network size decreases.
\item The mean distance between countries and the corresponding std are the largest for the UMLP networks and the smallest for the corresponding LMST networks.
\item The analysis shows the existence of a globalization process since  1960 till 1970 and its stabilisation thereafter, followed by a destabilisation after 2000 as observed in the decrease of the network size. 
\item The observation of the globalization process does not depend on the type of network constructed.
\end{itemize}

A word is in order concerning the time lag $\tau$ which could be introduced in the analysis. See Eq. (4). The time lag leads to an asymmetry in the correlation between the fluctuations. This induces us to suggest further studies which could lead to conclude on deducing a set of leaders and followers. 
We have observed in another though related analysis [Ausloos \& Mi\'skiewicz, 2008] that for increasing time lag the mean distance between network nodes increases, whence magnifying details of evolution.

Finally let us stress the interest of studying graphs, in particular to derive weighted networks such as in this paper, in order to have some comparative data organisation coherence.\footnote{This paper seems to  reproduce some considerations from [Ausloos \& Mi\'skiewicz, 2008]. We are aware that the presentation of results based on the same similar data set, through two different  mappings might induce some confusion in the reader. Our considerations in this paper corresponds indeed to taking $q=1$ in [Ausloos \& Mi\'skiewicz, 2008].  It should be emphasized that similar conclusions are obtained indeed, and listed in a similar way, but both works should be placed facing each other, adding to our comprehension of the time series mapping, network construction for analysis, and econophysics conclusions, rather than excluding one or another.}

\bigskip

\noindent {\bf Acknowledgments} \smallskip

\noindent Thanks to the organizers of Medyfinol'08 in Punta del Este, in particular C. Masoller. Let A. Proto be congratulated  
for suggesting MA participation at such a meeting, and F. Redelico for his help before, after and throughout the affair. 

\bigskip

\noindent {\bf References} \smallskip

\noindent Ausloos, M. \& Gligor, M. [2007] ``Cluster structure of EU-15 countries derived from the correlation
  matrix analysis of macroeconomic indices fluctuations,'' {\it Eur. Phys. J B} {\bf 57}, 139--146.

\noindent Ausloos, M. \& Gligor, M. [2008] ``Cluster expansion method for evolving weighted networks having
  vector-like nodes,'' {\it Acta Phys. Polon. A} {\bf 114}, 491--499.

\noindent Ausloos, M. \&  Lambiotte, R. [2007] ``Clusters or networks of economies? A macroeconomy study through GDP
  fluctuation correlations,'' {\it Physica A} {\bf 382}, 16--21.

\noindent Ausloos, M. \& Mi\'skiewicz, J. [2008]``Introducing the q-Theil index,'' {\it Braz. J. Phys.}, submitted for publications.

\noindent Gligor, M. \& Ausloos, M. [2008a] ``Clusters in weighted macroeconomic networks: the EU case.
  Introducing the overlapping index of GDP/capita fluctuation correlations,'' {\it Eur. Phys. J. B}, {\bf 63}, 533--539. 

\noindent Gligor, M. \& Ausloos, M. [2008b] ``Convergence and cluster structures in EU area according to
  fluctuations in macroeconomic indices,'' {\it J. Econ. Integration} {\bf 23}, 297--330.

\noindent Mi\'skiewicz, J. [2008] ``Globalization-Entropy unification through the Theil index,'' {\it Physica A} {\bf 387}, 6595--6604.

\noindent Mi\'skiewicz, J. \& Ausloos, J. [2005] ``Correlations between the most developed (G7) countries. A moving
  average window size optimisation,'' {\it Acta Phys. Pol. B}  {\bf 36}, 2477--2486.

\noindent Mi\'skiewicz, J. \& Ausloos, M. [2006] ``An {A}ttempt to observe {E}conomy {G}lobalization: {T}he {C}ross
  {C}orrelation {D}istance {E}volution of the {T}op 19 {GDP}'s,'' {\it Int. J. Mod. Phys. C} {\bf 17}, 317--331.

\noindent Mi\'skiewicz, J. \& Ausloos, M. [2008] ``Correlation measure to detect time series distances, whence economy
  globalization,'' {\it Physica A} {\bf 387}, 6584--6594.

\noindent Yang, Y. \& Yang, H. [2008] ``Complex network-based time series analysis,'' {\it Physica A} {\bf 387}, 1381--1386.

\end{document}